\documentclass[12pt]{article}

\usepackage[utf8]{inputenc}
\usepackage[T1]{fontenc}
\usepackage{lmodern}
\usepackage[letterpaper,margin=1in]{geometry}
\usepackage{setspace}
\usepackage{graphicx}
\usepackage{booktabs}
\usepackage{multirow}
\usepackage{array}
\usepackage{amsmath}
\usepackage{amssymb}
\usepackage{xcolor}
\usepackage{authblk}
\usepackage[numbers,sort&compress]{natbib}
\usepackage[hidelinks]{hyperref}
\usepackage{caption}

\newcommand{\Dp}{\ensuremath{D_p}}
\newcommand{\Dt}{\ensuremath{D_t}}
\newcommand{\fp}{\ensuremath{f_p}}
\newcommand{\ADC}{\mbox{ADC}}

\newcommand{\figorplaceholder}[2][\linewidth]{%
  \IfFileExists{#2}%
    {\includegraphics[width=#1]{#2}}%
    {\setlength{\fboxsep}{10pt}%
     \fbox{\parbox[c][0.26\textheight][c]{0.95\linewidth}{%
       \centering\small\nolinkurl{#2}\\[8pt]%
       \itshape Figure placeholder \nolinkurl{#2}.}}}%
}

\title{\bfseries An integrated diffusion-weighted imaging processing and interpretation platform for MR-guided radiotherapy}

\author[1,2]{Yunxiang Li}
\author[1]{Yan Dai}
\author[1]{Yen-Peng Liao}
\author[1]{Jie Deng}
\author[1]{Jill B. De Vis}
\author[1,*]{You Zhang}
\affil[1]{Department of Radiation Oncology, University of Texas Southwestern Medical Center, Dallas, Texas, USA}
\affil[2]{Department of Radiation Oncology, University of California San Francisco, San Francisco, California, USA}
\affil[*]{Corresponding author: You Zhang (\texttt{you.zhang@utsouthwestern.edu})}

\date{}

\begin{document}

\maketitle

\thispagestyle{empty}

\begin{onehalfspacing}

\section*{Abstract}
\noindent\textbf{Background:} Magnetic resonance imaging-guided linear accelerators (MR-Linacs) allow diffusion-weighted imaging (DWI) to be acquired at every treatment fraction, but converting these low-signal-to-noise-ratio acquisitions into clinical decisions requires both reliable quantitative processing and an interpretation that reconciles a scattered and often contradictory literature.

\noindent\textbf{Purpose:} To describe and evaluate an integrated, web-based platform that carries raw MR-Linac DWI to a structured, literature-grounded clinical interpretation, and to assess its retrieval-augmented generation (RAG) interpretation module by independent expert rating.

\noindent\textbf{Methods:} The platform couples a deep-learning processing pipeline, comprising distortion correction, denoising, and intravoxel incoherent motion (IVIM)/apparent diffusion coefficient (ADC) fitting, with longitudinal region-of-interest analysis and a RAG interpretation agent. The agent reasons over a two-layer knowledge base of curated publications (a structured catalog index plus line-indexed full text), delegates arithmetic to deterministic tools, and is designed to trace each statement to a source document, section, and line range. One medical physicist and one physician independently rated the agent's reports for nine longitudinal glioblastoma cases on a 1--5 scale across three metrics: clinical-reasoning soundness, literature-citation quality, and overall clinical utility.

\noindent\textbf{Results:} Across 54 ratings, the pooled mean was $4.65\pm0.80$, with 93\% of ratings $\ge 4$; Metric means were 4.6 (reasoning), 4.5 (citation), and 4.8 (utility), and raters agreed within one point on 85\% of paired ratings.

\noindent\textbf{Conclusions:} A single platform can integrate MR-Linac DWI post-processing with traceable, expert-evaluated clinical interpretation, while highlighting the safeguards needed to verify LLM-generated reasoning in radiation oncology.

\vspace{0.8em}
\noindent\textbf{Keywords:} diffusion-weighted imaging; intravoxel incoherent motion; MR-guided radiotherapy; large language models; retrieval-augmented generation; glioblastoma

\end{onehalfspacing}

\doublespacing

\section{Introduction}
\label{sec:intro}

Magnetic resonance imaging-guided radiotherapy (MRgRT) delivered on a magnetic resonance imaging-guided linear accelerator (MR-Linac) has changed not only how radiation is delivered but also what can be observed during a treatment course. Because the patient is imaged on the treatment table at each fraction, functional sequences such as diffusion-weighted imaging (DWI) can be concurrently acquired with anatomical sequences, rather than at the one or two isolated diagnostic time points typical of conventional follow-up~\cite{boeke2025,lawrence2021}. This per-fraction capability provides a longitudinal sampling density that diagnostic MRI cannot match and allows the tumor to be monitored throughout the treatment course. DWI and its intravoxel incoherent motion (IVIM) modeling are attractive functional biomarkers in this setting: the apparent diffusion coefficient (ADC) reflects tissue cellularity, while the IVIM model separates the signal into true tissue diffusion (\Dt), pseudo-diffusion (\Dp), and the perfusion fraction (\fp), thereby disentangling cellular from microvascular contributions~\cite{lebihan1988,iima2016,lebihan2019}. Sampling these parameters fraction by fraction offers, in principle, an early and biologically specific measure of how a tumor is responding to therapy. In this study, we described a platform specifically developed for this acquisition regime.

Both the opportunity and the difficulty are greatest in glioblastoma (GBM) management. Despite maximal chemoradiation, median survival remains approximately 15 months~\cite{stupp2005}, and post-treatment management is confounded by pseudoprogression, in which treatment-induced blood-brain barrier disruption and inflammation mimic true tumor recurrence on conventional contrast-enhanced MRI~\cite{brandsma2008,brandsma2009}. This phenomenon is common in a substantial fraction of patients overall and at especially high rates in MGMT-methylated tumors~\cite{brandes2008mgmt,kucharczyk2016}. Morphological response criteria cannot reliably separate these two biologically distinct processes~\cite{wen2010rano}, and confirmation typically requires months of follow-up, during which patients with genuine progression may be undertreated and miss the survival benefit of timely second-line therapy~\cite{nava2014}. In a meta-analysis of 24 studies (900 patients), DWI distinguished true progression from pseudoprogression with a pooled sensitivity of 0.88 and specificity of 0.85~\cite{tsakiris2020}. ADC alone, however, conflates diffusion and perfusion into a single quantity, limiting its ability to distinguish the complex biological processes in GBM, where VEGF-driven angiogenesis~\cite{plate1992} coexists with a heterogeneous microenvironment of viable tumor, necrosis, and an angiogenic rim ~\cite{hambardzumyan2015}. The multi-compartment IVIM model is better matched to this biology and has shown diagnostic and prognostic value in glioma, including tumor grading and treatment-response assessment~\cite{maralani2021,puig2016,liao2023,luo2021}.

Realizing the potential of DWI on an MR-Linac for response monitoring requires solving two separate problems. The first is technical. MR-Linac DWI is acquired at lower SNR than diagnostic DWI due to machine constraints, and clinically pragmatic single-shot echo-planar acquisition introduces geometric distortions; both effects destabilize quantitative fitting, especially of the perfusion-sensitive \Dp{} and \fp{} parameters~\cite{lebihan2019,henriksen2022}. We previously developed a dedicated processing chain for this low-SNR regime: landmark-matched B-spline implicit neural representation (INR) for distortion correction~\cite{li2026lmbs}, band-limited INR for denoising~\cite{li2026blinr}, and an INR-based IVIM parameter estimator that recovers reproducible longitudinal parameter maps~\cite{li2026accurate}. The second problem is interpretive and has received comparatively less attention. Even when reliable parameters are available, turning \Dp, \Dt, and \fp{} trajectories into a clinical assessment demands synthesizing evidence that is scattered across previous literature and often contradictory, because reported thresholds and parameter-outcome relationships vary with field strength, $b$-value protocol, vendor, tumor location, and treatment regimen~\cite{henriksen2022,mesny2024}. The three parameters may also move in opposite directions in the same patient (for example, \Dt{} decreasing while \fp{} rises), and no validated framework exists to resolve such multi-parameter discordance~\cite{lebihan2019}. On many occasions, the barrier to adopting IVIM for treatment monitoring outside specialized centers is this knowledge-synthesis burden rather than a shortage of data.

Software support exists for part but not the whole of the workflow. For instance, general DICOM-management tools can streamline data organization and conversion for downstream analysis~\cite{maniscalco2026mdh}, and several quantitative DWI methods have been validated on the MR-Linac~\cite{habrich2022,kooreman2019}. However, an integrated path progressing from raw MR-Linac DWI to an interpretation that a clinician can act on and audit is still missing. On the interpretation side, large language models (LLMs) coupled with retrieval-augmented generation (RAG) offer a natural technical approach. Rather than relying on a model's parametric memory, which risks fluent but unsupported conclusions, RAG requires the model to retrieve published evidence before it reasons~\cite{lewis2020rag,shuster2021}, mirroring how a clinician consults the literature before forming an assessment. Agentic LLM systems built on this principle have begun to deliver strong diagnostic performance in other medical domains while keeping their reasoning traceable: DeepRare, an agentic rare-disease system evaluated across nine datasets spanning 2{,}919 diseases, achieved an average Recall@1 of 57.18\% in human-phenotype-ontology-based diagnostic tasks, outperforming the next best method by 23.79\%, and expert review confirmed the validity of 95.4\% of its reasoning chains~\cite{zhao2026deeprare}.

In this work, we develop and evaluate an integrated, web-based platform that aims to address both the technical and interpretive challenges of DWI's clinical application. The platform carries MR-Linac DWI from raw DICOM through distortion correction, denoising, IVIM/ADC fitting, and registration, to longitudinal, region-of-interest (ROI)-based parameter trajectories, and then submits those trajectories to a RAG-based clinical interpretation agent that returns a structured report in which each statement is anchored on a specific passage of the source literature. We first present the system architecture, its data connectivity, the deep-learning processing components, and the design of the interpretation agent. We then report an evaluation of the agent's reports on nine longitudinal GBM cases, independently rated by a medical physicist and a physician across reasoning, citation, and utility metrics. Rather than proposing a new imaging algorithm, this work demonstrates that processing and interpretation can be unified in one auditable clinical tool, and reports where expert reviewers found the tool trustworthy and where it needs further improvement.

\section{Methods}
\label{sec:methods}

\subsection{System overview and architecture}
\label{sec:overview}

The platform is implemented as a single-page web application backed by a Python server. The backend uses the Flask framework with Flask-SocketIO for bidirectional, real-time communication, so that long-running processing jobs report progress to the browser as they execute \cite{grinberg2014flask}. The front end is an HTML/JavaScript client. All image processing runs locally on institutional hardware with GPU acceleration, and no image data leaves the local environment. Only the derived numerical summaries used by the interpretation agent are sent to the HIPAA-compliant language-model deployment described in Section~\ref{sec:agent}. The deep-learning components are built on PyTorch~\cite{paszke2019pytorch}; medical-image input/output and geometry are handled by pydicom~\cite{mason2011pydicom}, nibabel, and SimpleITK~\cite{lowekamp2013simpleitk}, with additional transform modules from MONAI~\cite{cardoso2022monai}. Table~\ref{tab:software} summarizes the principal software components and their roles.

\begin{figure}[htbp]
\centering
\figorplaceholder{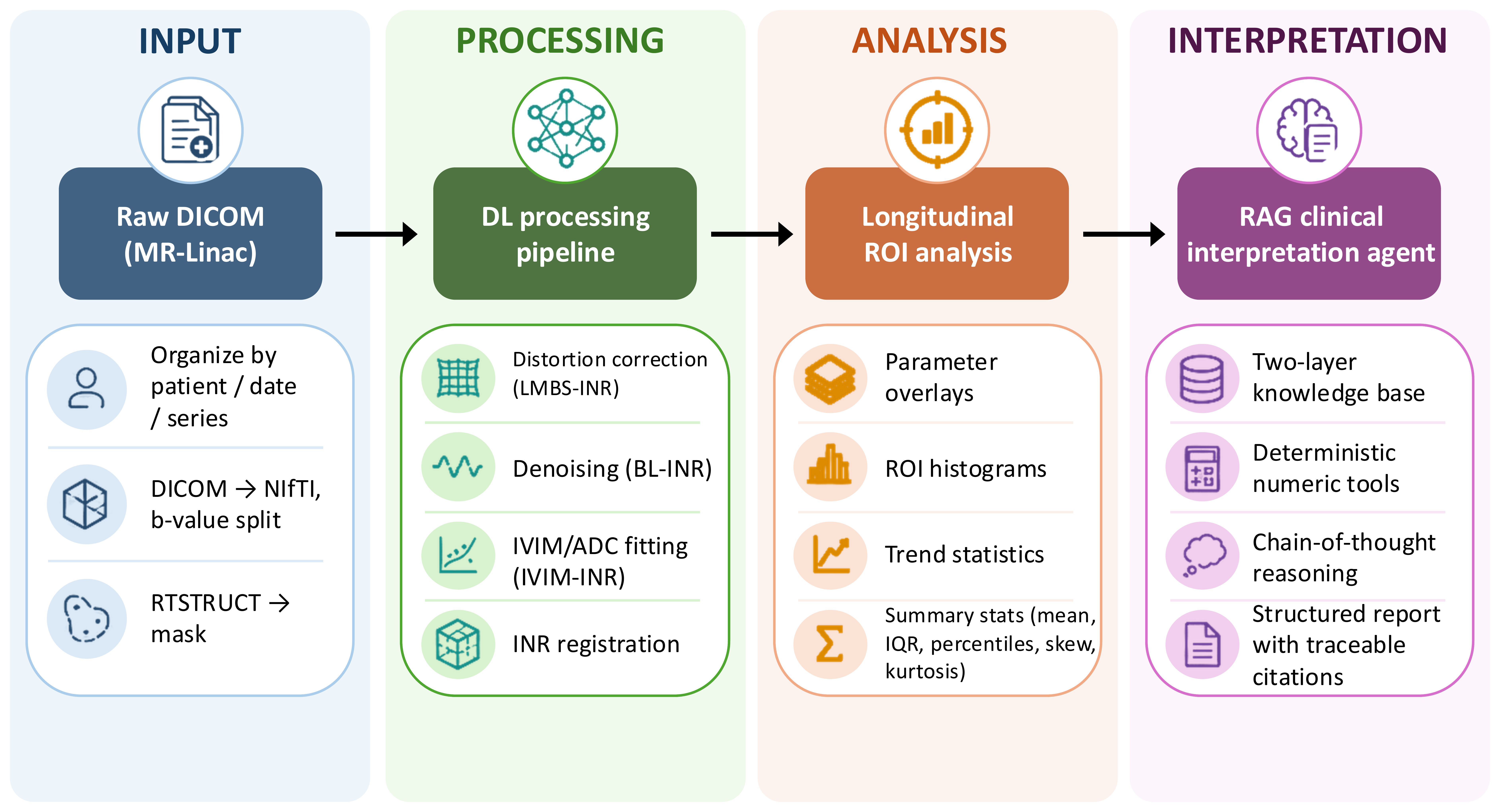}
\caption{End-to-end architecture and data flow of the platform. Raw MR-Linac DICOM data are organized and converted to NIfTI (with $b$-value separation for DWI and mask extraction for segmentations), processed through a deep-learning pipeline into IVIM/ADC parameter maps, analyzed longitudinally over regions of interest, and finally interpreted by a retrieval-augmented generation (RAG) agent that is designed to trace evidence-supported statements to line-indexed source literature passages. Processed maps can be exported back to DICOM.}
\label{fig:arch}
\end{figure}

The application is organized around four user-facing capabilities, exposed from a common landing page: an ADC\,\&\,IVIM processing pipeline; an MRI analysis module for longitudinal, ROI-based quantification and visualization; a clinical-interpretation agent; and a DICOM export utility that returns processed maps to clinical systems. Internally, processing is expressed as a numbered sequence of steps that share a common on-disk data hierarchy keyed by patient identifier and study date, so that intermediate results are inspectable and any step can be re-run independently. The overall data flow, from raw DICOM through quantitative maps to a traceable interpretation, is shown in Figure~\ref{fig:arch}. Representative views of the processing, analysis, and interpretation interfaces are presented alongside the corresponding modules below. A defining feature of the platform is that these stages form a single application rather than separate tools: a study can be carried from DICOM import to a clinical report without leaving the browser or writing any code.

\begin{table}[t]
\centering
\caption{Principal software components of the platform and their roles. Versions reflect the validated deployment configuration.}
\label{tab:software}
\small
\begin{tabular}{@{}l l p{0.42\linewidth}@{}}
\toprule
\textbf{Component} & \textbf{Version} & \textbf{Role in the platform} \\
\midrule
Python                & 3.10       & Implementation language \\
Flask / Flask-SocketIO & 2.3 / 5.5 & Web server and real-time progress streaming \\
PyTorch               & 2.6 (CUDA 12.6) & Deep-learning models (distortion, denoising, fitting) \\
pydicom               & 3.0.1      & DICOM reading and metadata extraction \\
nibabel               & 5.3.2      & NIfTI input/output \\
SimpleITK             & 2.4.1      & Reorientation, resampling, geometry \\
MONAI                 & 1.4.0      & Medical-image transforms \\
NumPy / SciPy         & 1.26 / 1.15 & Numerical computation and optimization \\
LangChain / LangGraph & n/a        & Agent orchestration and tool calling \\
Azure OpenAI (GPT-5.1) & n/a       & Large-language-model reasoning (HIPAA-compliant) \\
\bottomrule
\end{tabular}
\end{table}

\subsection{Data connectivity and DICOM handling}
\label{sec:data}

Clinical interoperability is handled entirely through DICOM. Raw exports are first organized into a deterministic hierarchy (\texttt{patient/date/series}). Anatomical sequences (T2-weighted, FLAIR, T1-weighted) are converted to NIfTI directly; DWI series are parsed for their diffusion $b$-value and written as one volume per $b$-value, rendering a multi-$b$-value stack required for IVIM/ADC fitting. During conversion, slices are sorted by their physical position, and volumes are reoriented to a consistent voxel grid, preserving the affine geometry needed for later registration. Segmentation objects (DICOM-SEG and RT-STRUCT) are converted to binary mask volumes that are spatially matched to their referenced anatomical image through the DICOM frame-of-reference metadata, which makes clinical target volumes (e.g., GTV, CTV) and organs at risk directly available as ROIs for analysis. After processing, parameter maps can be written back to DICOM, preserving patient and study metadata so that quantitative outputs can be reviewed alongside the planning images in clinical systems.

\subsection{Image-processing pipeline}
\label{sec:pipeline}

\begin{figure}[htbp]
\centering
\figorplaceholder{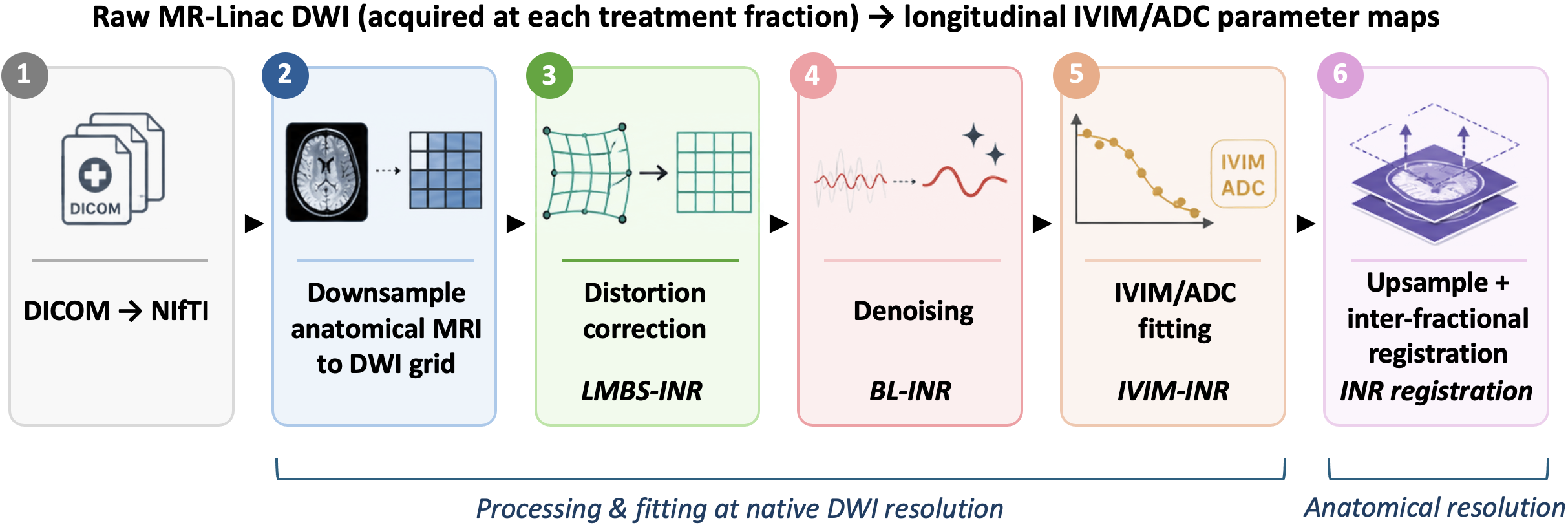}
\caption{The DWI/IVIM processing pipeline. Raw multi-$b$-value MR-Linac DWI, acquired at each treatment fraction, is processed through six steps; quantitative fitting is performed at native DWI resolution (steps 2--5) and only the final maps are resampled to anatomical resolution before inter-fractional registration (step 6).}
\label{fig:pipeline}
\end{figure}

The core ADC\,\&\,IVIM pipeline executes as an ordered sequence of steps
(Figure~\ref{fig:pipeline}). Quantitative processing is performed at the native DWI resolution, with only the final parameter maps resampled to match the higher-resolution anatomical images. This substantially reduces fitting time compared with fitting upsampled DWI volumes while ensuring that parameter estimation is based on the originally acquired signal. The deep-learning components, summarized below, have been described and validated in detail elsewhere and are used here as integrated modules.

\paragraph{Anatomical downsampling.} The high-resolution anatomical reference is resampled to the DWI grid to serve as a structural prior for distortion correction and denoising.

\paragraph{Distortion correction (LMBS-INR).} Geometric distortion from $B_0$ inhomogeneity is corrected by deformably registering the DWI volume to the anatomical reference. Our approach combines cross-modality landmark matching from a modality-invariant image-matching model~\cite{ren2025minima} with a B-spline implicit neural representation of the deformation field, overcoming the local optima of conventional intensity-only optimization while enforcing smooth, anatomically plausible deformations. On brain GBM data, the method achieved a mean Dice coefficient of 0.919, exceeding established diffeomorphic and learning-based baselines, which ensures that ROI-based parameter measurements correspond to consistent anatomy across fractions~\cite{li2026lmbs}.

\paragraph{Denoising (BL-INR).} The low SNR of MR-Linac DWI is addressed with a self-supervised, band-limited implicit neural representation that restricts the network's learnable frequency content to suppress high-frequency noise, while using the multi-$b$-value signal-decay relationship and cross-modality structural consistency as physics-informed constraints. The method requires no paired training data and, on clinical DWI, reduced the noise standard deviation to 40\% of its original level while preserving the signal mean, achieving a structural similarity index of 0.933~\cite{li2026blinr}.

\paragraph{IVIM/ADC parameter fitting (IVIM-INR).} Voxel-wise IVIM fitting is highly sensitive to noise, particularly for \fp{} and \Dp. Treating each voxel independently ignores spatial continuity and yields noisy maps with poor longitudinal consistency. We instead frame fitting as spatial-function learning: a periodic-activation (SIREN) implicit neural representation~\cite{sitzmann2020siren} models each parameter map as a continuous function of position, so that structural context and inter-voxel correlation are leveraged intrinsically~\cite{li2026accurate}. The network fits the IVIM bi-exponential model (Eq.~\ref{eq:ivim}),
\begin{equation}
\frac{S(b)}{S_0} = \fp\, e^{-b\Dp} + (1-\fp)\, e^{-b\Dt},
\label{eq:ivim}
\end{equation}
and, on the low-$b$/high-$b$ partition, the mono-exponential ADC model (Eq.~\ref{eq:adc}),
\begin{equation}
\frac{S(b)}{S_0} = e^{-b\,\ADC},
\label{eq:adc}
\end{equation}
with physically motivated bounds ($\Dp\!\in\![3.5\times10^{-3},\,0.1]$, $\Dt\!\in\![1\times10^{-5},\,3\times10^{-3}]$, $\fp\!\in\![0,\,0.99]$, diffusivities in $\mathrm{mm}^2/\mathrm{s}$). Across multi-time-point MR-Linac data, this approach improved longitudinal reproducibility for all three IVIM parameters over least-squares and recent deep-learning baselines (for example, raising the \fp{} intraclass correlation coefficient from 0.29 to 0.56), which is essential for detecting small but clinically meaningful changes over the course of treatment.~\cite{li2026accurate}.

\paragraph{Registration and resampling.} The pipeline includes INR-based rigid and deformable registration. After fitting, native-resolution DWI and parameter maps are upsampled to the anatomical reference, and inter-fractional deformable registration brings all fractions into a common anatomical space, so that a fixed ROI can be propagated across the treatment course for longitudinal comparison. For these inter-fractional alignments, a mask-based rigid pre-alignment (excluding tumor-bearing regions so that disease change does not bias the transform) is performed before the deformable refinement.

\subsection{Longitudinal analysis and visualization}
\label{sec:analysis}

\begin{figure}[htbp]
\centering
\figorplaceholder[0.8\linewidth]{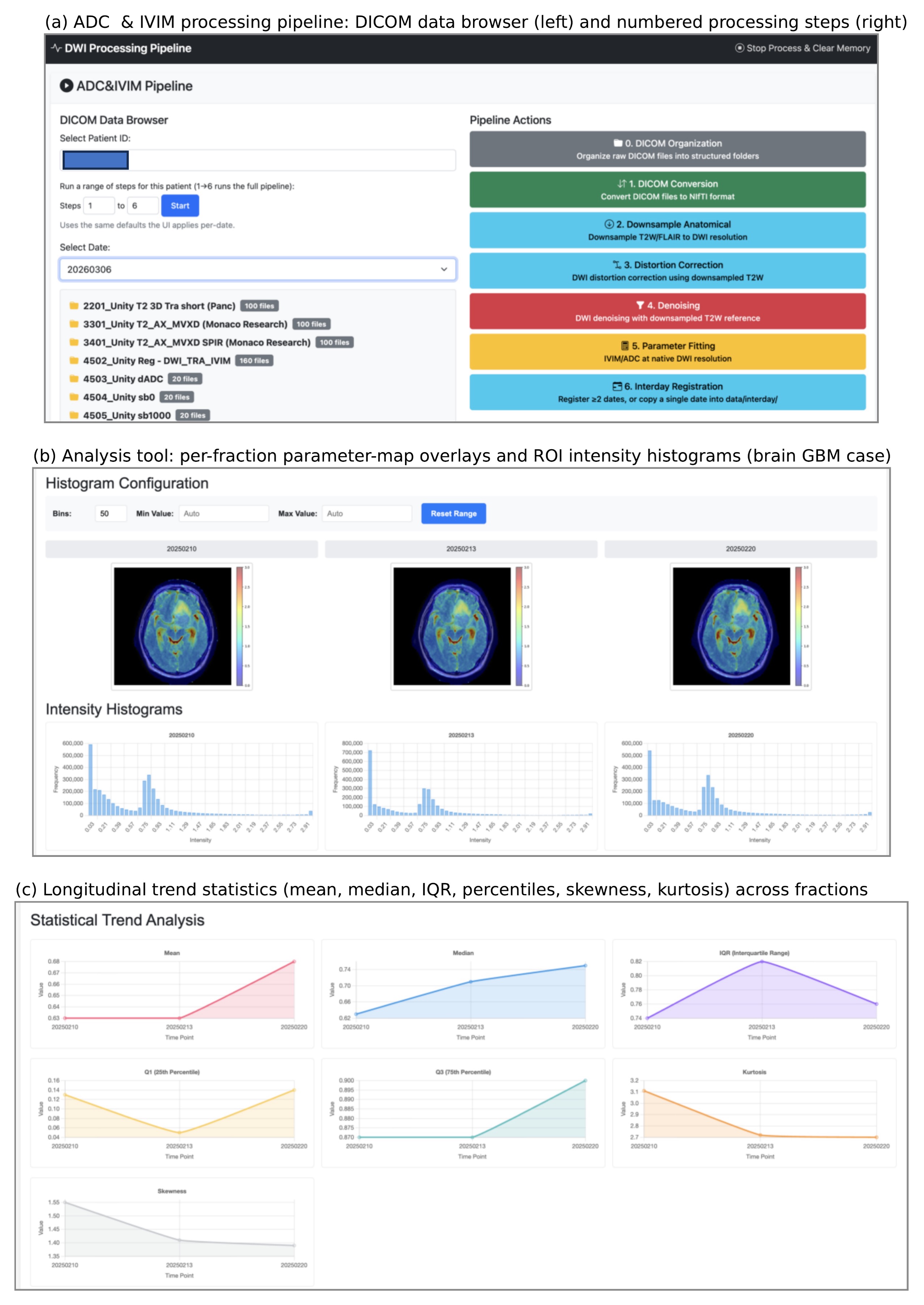}
\caption{The web platform interface, with a brain GBM case shown. (a) The ADC \& IVIM processing pipeline, with the DICOM data browser (patient/date/series, left) and the numbered processing steps (right). (b) The analysis tool, showing per-fraction parameter-map overlays on the anatomical image and the corresponding ROI intensity histograms. (c) Longitudinal trend statistics (mean, median, interquartile range, percentiles, skewness, and kurtosis) computed within a fixed ROI across treatment fractions.}
\label{fig:ui}
\end{figure}

The analysis module turns registered parameter maps into quantitative trajectories that drive interpretation. For a user-selected ROI imported from RT-STRUCT, the module overlays each parameter map on the anatomical image at every time point, computes ROI-based intensity histograms (Figure~\ref{fig:ui}b), and reports a panel of longitudinal summary statistics (mean, median, interquartile range, selected percentiles, skewness, and kurtosis) as functions of fraction date (Figure~\ref{fig:ui}c). A fixed first-fraction ROI can be applied to all time points, or date-specific ROIs can be used. The resulting per-parameter trajectories and their percentage changes constitute the structured numerical input to the interpretation agent, ensuring that the agent reasons over the same auditable quantities a physicist/physician would inspect manually.

\begin{figure}[htbp]
\centering
\figorplaceholder{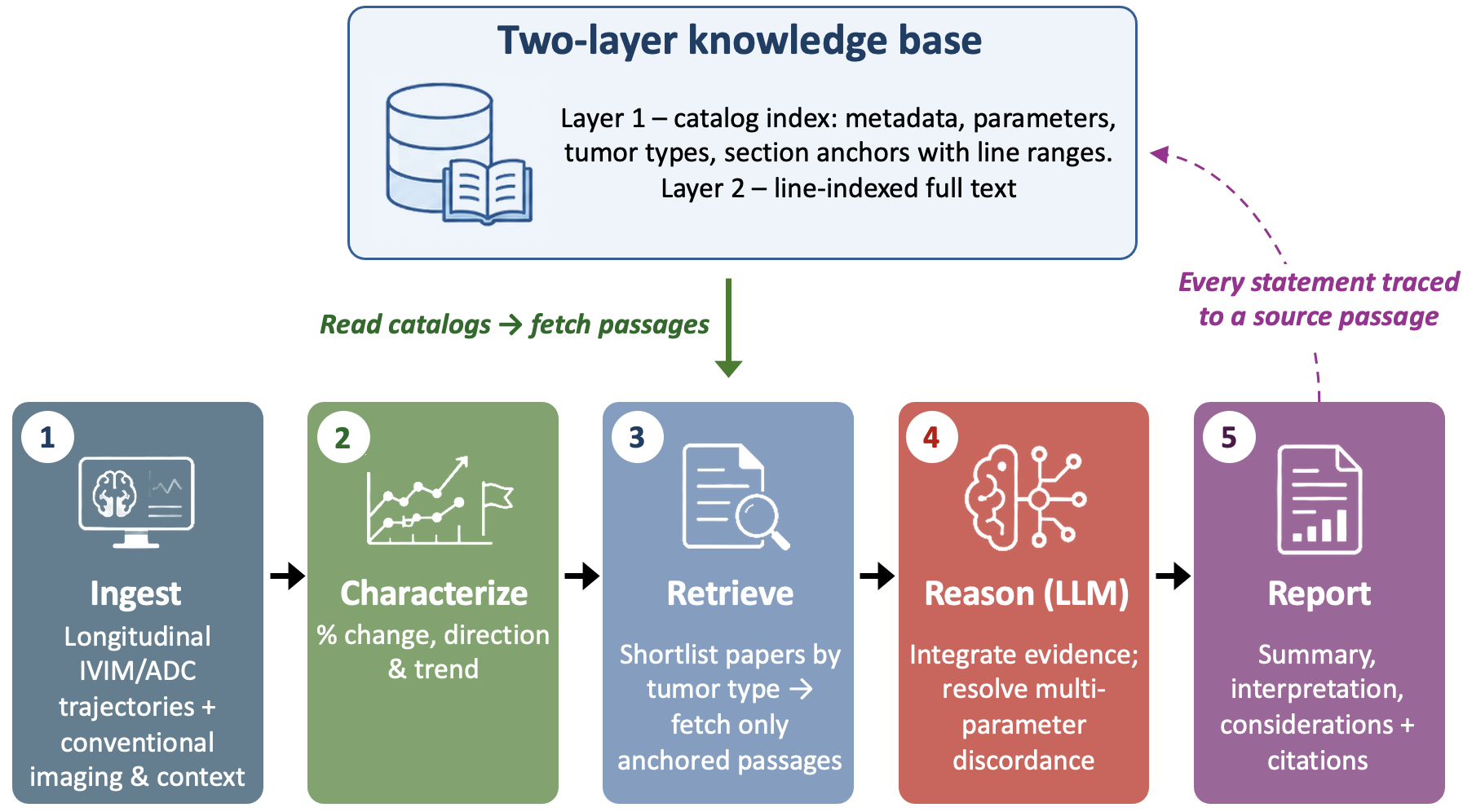}
\caption{The retrieval-augmented interpretation agent. Longitudinal IVIM/ADC trajectories, together with the conventional imaging context, are ingested (stage 1) and characterized (stage 2). Catalog-guided retrieval then shortlists tumor-type- and parameter-matched publications from the two-layer knowledge base and fetches only the anchored passages (stage 3). The LLM integrates the evidence and resolves multi-parameter discordance (stage 4), and a structured report is produced in which evidence-supported statements are traced back to a source document, section, and line range (stage 5).}
\label{fig:agent}
\end{figure}

\subsection{Retrieval-augmented clinical-interpretation agent}
\label{sec:agent}

The interpretation agent is the platform's key component. It is built with the LangChain/LangGraph orchestration framework~\cite{chase2022langchain} and uses a GPT-5-class LLM (GPT-5.1) served through our institution's HIPAA-compliant Microsoft Azure OpenAI deployment. Three design choices (Figure~\ref{fig:agent}) constrain the agent to ground its output in the retrieved literature rather than generate free-form text: a structured knowledge base, two-stage retrieval, and tool-assisted calculations with passage-level citation.

\begin{figure}[htbp]
\centering
\figorplaceholder[\linewidth]{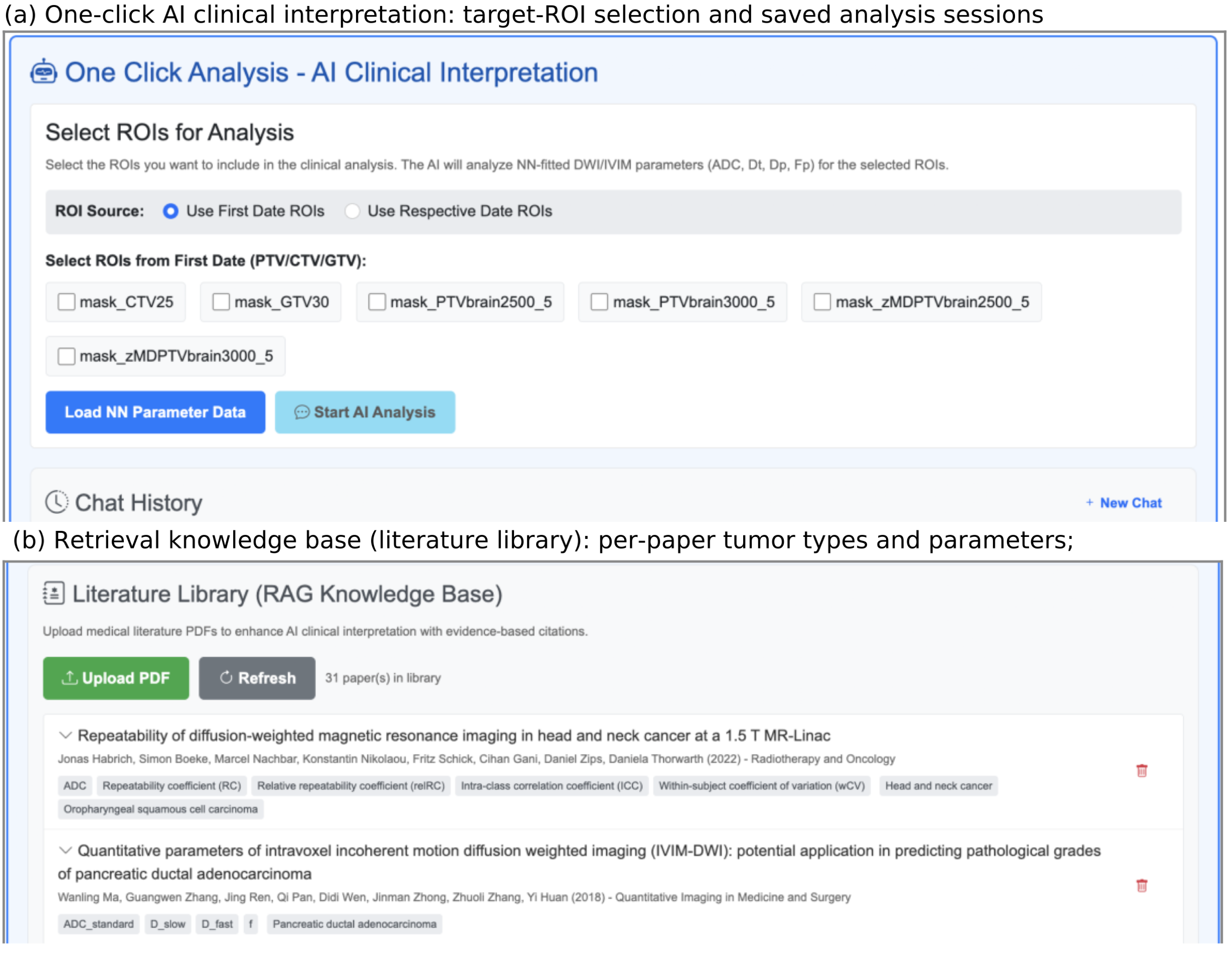}
\caption{The clinical-interpretation interface. (a) The one-click AI clinical-interpretation view, where the user selects the target ROIs (e.g., GTV/CTV/PTV imported from RT-STRUCT) and launches an analysis. Prior analysis sessions are retained for review. (b) The retrieval knowledge base (``literature library''), which lists each indexed publication with its tumor types and reported parameters and accepts new papers by PDF upload.}
\label{fig:agentui}
\end{figure}

\paragraph{Two-layer knowledge base.} At the time of this study, the evidence base comprised 28 peer-reviewed publications on DWI/IVIM (and related quantitative MRI) in oncology and MRgRT, curated to span the tumor types and parameters relevant to the platform (Table~\ref{tab:kb}). Each publication is stored in two layers. Layer~1 is a structured catalog: a machine-readable record of the paper's metadata (title, authors, journal, year, DOI), the tumor types and treatments studied, the IVIM/ADC and other parameters it reports, and a section-level content map in which every section is summarized and anchored to an explicit line range in the source text. Layer~2 is the paper's full text, stored as line-indexed plain text so that any anchored passage can be retrieved word by word on demand. This mirrors how humans usually read---scanning a table of contents before turning to the relevant section. The strategy also allows the agent to reason over a concise overview of the entire corpus before accessing any full-text documents. New papers can be continuously added through the interface by uploading a PDF (Figure~\ref{fig:agentui}b), which is parsed into both layers automatically.

\paragraph{Two-stage retrieval and tool-augmented reasoning.} The user selects the target ROIs and launches the analysis from the interpretation interface (Figure~\ref{fig:agentui}a). Given a patient's longitudinal parameter trajectories, the agent then proceeds in stages. It first characterizes the parameter changes (percentage change, direction, and trend for each of ADC, \Dp, \Dt, and \fp), delegating these computations to deterministic Python tools rather than performing calculations itself, which removes a well-known LLM failure mode. It then performs catalog-guided retrieval: reading the Layer-1 catalogs, it shortlists publications whose tumor type, parameter scope, and reported change directions match the patient's pattern, explicitly filtering for tumor-type compatibility (for example, restricting to GBM/brain studies for a brain case), and retrieves only the specific Layer-2 passages indicated by the matching catalog anchors. Finally, it integrates the tool-computed findings, user-provided prompt, and retrieved passages through chain-of-thought reasoning to identify multi-parameter discordances and resolve them by reference to biological mechanisms supported by the retrieved evidence.

\paragraph{Traceable report generation.} The agent returns a structured report with a parameter summary, an evidence-based interpretation, and clinical considerations. Citation is implemented through explicit passage tracking rather than relying on the LLM's recollection: each retrieved Layer-2 passage carries its source document, section, and line range, which are propagated to any statement derived from it, so that a reader can verify each claim against the exact lines of the cited paper (Figure~\ref{fig:agent}). This traceability makes the system auditable and is the key property that our expert evaluation was designed to assess.

\begin{table}[htbp]
\centering
\caption{Composition of the retrieval knowledge base (28 publications). Counts of tumor types and reported parameters exceed 28 because individual papers may span multiple categories.}
\label{tab:kb}
\small
\begin{tabular}{@{}l p{0.46\linewidth} l@{}}
\toprule
\textbf{Attribute} & \textbf{Categories (examples)} & \textbf{Coverage} \\
\midrule
Tumor type & Glioblastoma / glioma / brain metastases & predominant \\
           & Pancreatic adenocarcinoma; head and neck; prostate & secondary \\
Parameters & ADC & 28 papers \\
           & IVIM (\fp, \Dt, \Dp) & 13 papers \\
Setting    & MR-Linac / MR-guided radiotherapy; chemoradiation \\
Evidence   & Original studies, systematic reviews/meta-analyses, position statements & mixed \\
\bottomrule
\end{tabular}
\end{table}

\subsection{Expert evaluation of the interpretation agent}
\label{sec:evaldesign}

We evaluated the clinical quality of the agent's reports on nine longitudinal GBM cases whose MR-Linac studies were processed end-to-end through the platform. For each case, longitudinal DWI was processed into IVIM/ADC trajectories within the clinical gross tumor volume (GTV), and the agent generated a structured interpretation report. Two domain experts (one medical physicist and one physician with neuro-oncology imaging expertise) independently reviewed all nine reports.

Each report was rated on a 1--5 Likert scale along three pre-specified metrics defined in a shared rubric: (i) clinical-reasoning soundness: whether the reasoning is rigorous and correctly analyzes concordant and discordant multi-parameter relationships; (ii) literature-citation quality: whether citations are relevant, accurate, and traceable to a specific section and line range matching the case; and (iii) overall clinical utility: whether the report would aid clinical decision-making. Anchors were defined for each level (5 = excellent; 4 = good; 3 = acceptable; 2 = deficient; 1 = poor), and raters could add free-text comments. Ratings were collected independently, with no communication between the reviewers.

Given the small pilot-scale cohort, the analysis is descriptive. We report per-metric mean\,$\pm$\,standard deviation for each rater and pooled across raters, the proportion of ratings at or above each scale anchor, and, as a simple inter-rater agreement check, the rate of exact and within-one-point agreement and the mean absolute difference between raters across the 27 paired score cells (9 cases $\times$ 3 metrics). We avoid inferential agreement statistics (e.g., weighted kappa or intraclass correlation), which can be unstable and easily over-interpreted at this sample size, particularly when one rater’s scores exhibit near-zero variance. Free-text comments were reviewed qualitatively to identify recurring strengths and failure modes.

\section{Results}
\label{sec:results}

\subsection{Output of a representative case}
\label{sec:demo}

The platform processed MR-Linac studies end-to-end, from raw DICOM to registered, longitudinal IVIM/ADC maps and a clinical interpretation, within the local environment and without manual scripting. Native-resolution fitting kept per-study processing tractable, and the analysis module produced ROI-based parameter trajectories suitable both for direct review and for use as agent input.

\begin{figure}[htbp!]
\centering
\figorplaceholder[0.9\linewidth]{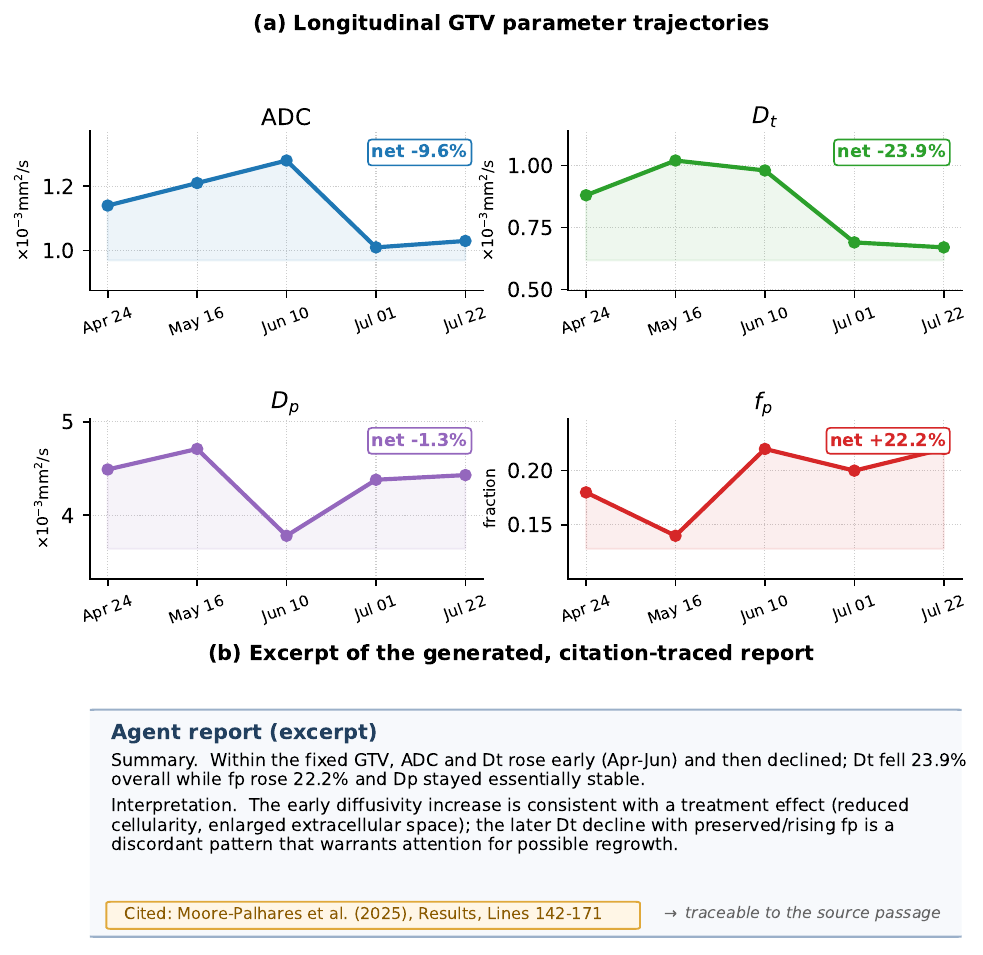}
\caption{A representative longitudinal GBM case. (a) GTV-mean trajectories of ADC, \Dt, \Dp, and \fp{} across five longitudinal time points, showing an early diffusivity rise followed by a late decline with preserved/rising perfusion fraction. (b) Excerpt of the agent's report, in which the interpretation is anchored to specific lines of matched source literature. This case received the maximum rating on all three dimensions from both reviewers.}
\label{fig:case}
\end{figure}

A representative case illustrates the output (Figure~\ref{fig:case}). For a GBM patient imaged at five longitudinal MR-Linac time points spanning the treatment course, the GTV delineated at the first time point was propagated to all subsequent scans. From the first to the last time point, ADC, \Dt{}, \Dp{}, and \fp{} changed by $-9.6\%$, $-23.9\%$,  $-1.3\%$, and  $+22.2\%$, respectively, with ADC and \Dt{} showing an early increase followed by a marked late decline. The agent's report first summarized these longitudinal changes and then interpreted them in phases. It attributed the early increase in diffusivity to an expected treatment response,  consistent with reduced cellularity and expansion of the extracellular space, while interpreting the subsequent decline in \Dt{} with preserved or rising \fp{} as a pattern warranting attention for possible tumor regrowth. The report also explicitly highlighted the discordance between decreasing diffusivity and stable or increasing perfusion. Each interpretive statement carried a citation to a specific section and line range of a matched literature source. For example, the treatment-response reading was anchored to longitudinal-ADC findings during chemoradiation, and the perfusion analysis to MRgRT brain-tumor literature~\cite{moorepalhares2025,lawrence2023,maziero2021,tsakiris2020}. Both reviewers rated this case 5/5/5 for the three metrics.

\begin{figure}[htbp!]
\centering
\figorplaceholder[0.85\linewidth]{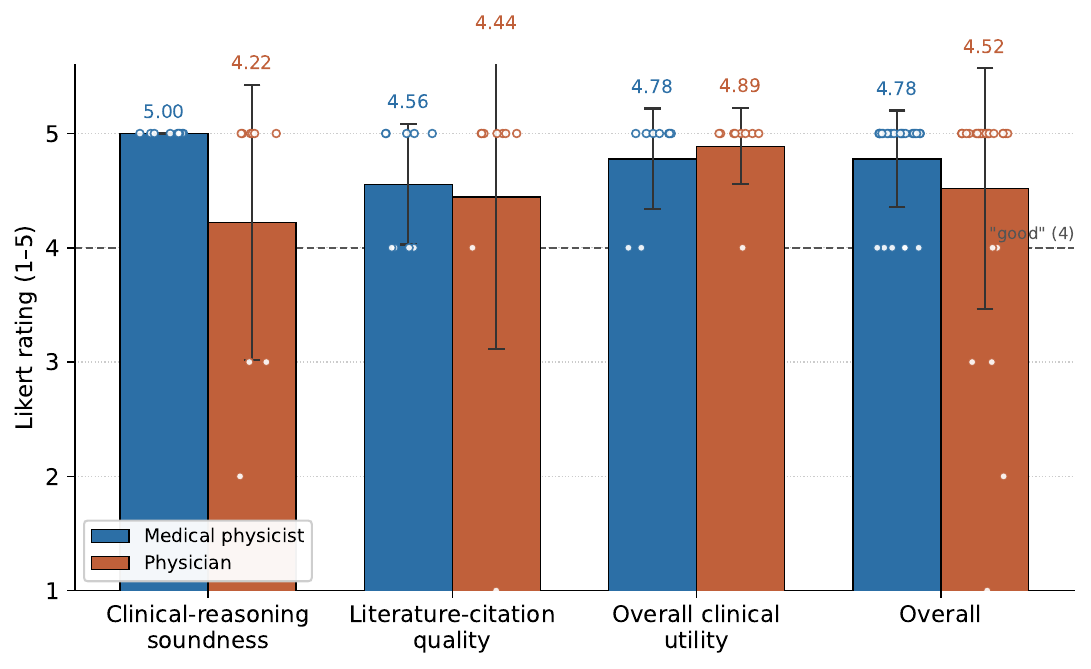}
\caption{Independent expert ratings of agent-generated reports by metric and rater (mean $\pm$ SD; 1--5 Likert). The dashed line marks the ``good'' scale anchor (4).Overall clinical utility received the highest and most consistent ratings from both reviewers. The physician's ratings showed greater variability, driven by a few atypical cases (reasoning scores of 2, 3, and 3 in three cases and one citation score of 1), whereas the physicist assigned a reasoning of 5 to all cases.}
\label{fig:eval}
\end{figure}

\subsection{Expert evaluation results}
\label{sec:evalresults}

Across the 54 ratings (9 cases $\times$ 3 metrics $\times$ 2 raters), the pooled mean score was $4.65\pm0.80$, and 93\% of ratings were $\ge 4$ (50/54), with 96\% $\ge 3$ (52/54). Pooled metric means were $4.61\pm0.92$ for clinical-reasoning soundness, $4.50\pm0.99$ for literature-citation quality, and $4.83\pm0.38$ for overall clinical utility (Table~\ref{tab:eval}, Figure~\ref{fig:eval}). The physicist's overall mean was $4.78\pm0.42$ (reasoning $5.00\pm0.00$, citation $4.56\pm0.53$, utility $4.78\pm0.44$) and the physician's was $4.52\pm1.05$ (reasoning $4.22\pm1.20$, citation $4.44\pm1.33$, utility $4.89\pm0.33$). The two reviewers agreed exactly on 59\% of the 27 paired ratings and within one point on 85\%, with a mean absolute difference of 0.63 points. The few larger gaps came from the physician's lower reasoning and citation scores on a small number of cases (detailed below); the remaining disagreements were one-point differences in both directions.

\begin{table}[t]
\centering
\caption{Expert evaluation of agent-generated reports on nine longitudinal GBM cases. Values are mean\,$\pm$\,standard deviation on a 1--5 Likert scale (higher is better). The pooled column combines both raters ($n=18$ per metric).}
\label{tab:eval}
\small
\begin{tabular}{@{}lccc@{}}
\toprule
\textbf{Dimension} & \textbf{Physicist} & \textbf{Physician} & \textbf{Pooled} \\
 & ($n=9$) & ($n=9$) & ($n=18$) \\
\midrule
Clinical-reasoning soundness   & $5.00\pm0.00$ & $4.22\pm1.20$ & $4.61\pm0.92$ \\
Literature-citation quality    & $4.56\pm0.53$ & $4.44\pm1.33$ & $4.50\pm0.99$ \\
Overall clinical utility       & $4.78\pm0.44$ & $4.89\pm0.33$ & $4.83\pm0.38$ \\
\midrule
\textbf{Overall}               & $4.78\pm0.42$ & $4.52\pm1.05$ & $4.65\pm0.80$ \\
\bottomrule
\end{tabular}
\end{table}

The free-text comments complemented the numerical scores. The physicist recorded numerical ratings without written remarks, while the physician annotated several cases. These annotations both affirmed the reports, noting for example that a borderline case (with changes between stable disease and possible recurrence) was nonetheless interpreted correctly, and localized failure modes that recurred across three reports. In one case (clinical-reasoning score 2, citation score 1), the agent cited additional GBM cohorts as supporting evidence without establishing their relevance to the index patient’s outcome and suggested a supplementary analysis (examining histograms or percentiles) without providing supporting citation. In two additional cases (clinical-reasoning score 3), the agent inferred the acquisition timeline (labeling scans as ``baseline,'' ``early treatment,'' or ``one month after radiation,'' and even estimating fraction numbers) from typical treatment schedules reported in the literature rather than from the study metadata. In one of these cases, the agent also introduced diffusivity measures derived from very high $b$-values (e.g., $b=2500$ and $3000\,\mathrm{s}/\mathrm{mm}^2$) without explaining how these measures should be interpreted or weighed. These failure modes reflect limitations in grounding rather than fluency, highlighting cases in which the platform's passage-tracing design did not fully achieve its intended purpose. We discuss the implications of these findings further in the Discussion.

\section{Discussion}
\label{sec:discussion}

This work combined two capabilities needed to reliably use per-fraction DWI on the MR-Linac: reliable processing, and interpretation of the resulting parameter trajectories against the published evidence. The platform described in this study unifies them in a single auditable tool, taking raw DICOM through validated deep-learning distortion correction, denoising, and IVIM/ADC fitting to longitudinal, ROI-based parameter maps, and then submitting those maps to a RAG agent that returns a structured, citation-traced interpretation. Domain experts judged these interpretations to be clinically usable. Pooled across 54 independent ratings, reports scored $4.65\pm0.80$ on a 1--5 scale, with overall clinical utility rated highest ($4.83\pm0.38$) and 93\% of all ratings at ``good'' or above.

Three design decisions likely contributed to these ratings. First, traceability: by linking citations directly to retrieved passages rather than relying on the LLM’s internal knowledge, the system enables reviewers to verify individual claims against the corresponding source text. Both experts viewed this capability favorably, and it represents an important distinction from a standalone LLM. However, citation received the lowest ratings in cases where claims were not adequately supported by matched evidence, indicating that traceability is only effective when grounding is consistently maintained. Second, separation of computation from reasoning: assigning percentage-change and trend calculations to deterministic tools minimizes arithmetic errors and ensures that subsequent interpretation is based on reproducible numerical results. Third, tumor-type-aware retrieval from a curated corpus: restricting retrieval to a vetted, line-indexed knowledge base and filtering evidence for case compatibility helps base the interpretation on clinically relevant literature. Together, these design choices address a key barrier to IVIM adoption: the burden of integrating and interpreting heterogeneous quantitative evidence \cite{henriksen2022,mesny2024},  rather than limitations in imaging performance or processing.

The evaluation also revealed the system's limits. The physician's lower scores corresponded to specific, identifiable failure modes rather than rating noise. The agent, when not given explicit acquisition metadata, inferred scan timing from typical treatment schedules in the literature; and in at least one case it drew on GBM cohorts that were not matched to the index case on the relevant outcome, and suggested an analysis without providing a supporting reference. These behaviors represent the residual risk that LLM-based interpreter carries, and point to implementable safeguards: (1). passing verified study metadata (true fraction numbers and dates) to the agent rather than letting it infer them; (2). constraining retrieval and cohort comparisons to studies with protocols and outcome definitions that are compatible with the case, while explicitly flagging any mismatches; (3). requiring that every analysis carry a citation or be labeled as general clinical reasoning; (4). restricting the report to a pre-specified parameter set, with any auxiliary metric (e.g., very-high-$b$ diffusivity) included only when its interpretive relevance and appropriate weighting are explicitly stated; and (5). reporting a confidence level that is reduced for atypical or under-specified cases rather than forcing a definitive interpretation. The fact that two reviewers, independently applying the same rubric, could pinpoint the specific statements in which these issues occurred provides evidence that the traceable-report design supports the intended level of auditability.

This platform differs in scope from existing tools. Research-oriented DICOM managers streamline data curation for downstream analysis~\cite{maniscalco2026mdh}, and a growing body of work has established the accuracy and repeatability of quantitative DWI on the MR-Linac~\cite{habrich2022,kooreman2019,lawrence2021}. The platform builds on this foundation by extending the workflow beyond quantification to interpretation and, consistent with broader shift toward agentic, evidence-grounded medical AI~\cite{zhao2026deeprare}, generates reports whose statements can be directly verified against their supporting sources. Although the validated clinical use case is longitudinal GBM, the architecture itself is organ-agnostic. The knowledge base already spans head-and-neck, prostate, and pancreatic disease, so the same processing-to-interpretation path can be potentially generalized by extending the corpus.

This study has several limitations. The evaluation is a single-institution, pilot-scale study with nine cases, two raters, and a single underlying LLM configuration, which supports feasibility and clinical-quality assessment but not deployment-ready generalization. The reports were rated for quality, reasoning, and utility, not against patient outcomes. In addition, we did not evaluate whether the agent's IVIM-based assessment distinguishes progression from pseudoprogression more accurately than an ADC-only baseline, because ground-truth outcome labels based on pathology or $\ge$6-month imaging follow-up were beyond the scope of this study.

\section{Conclusions}
\label{sec:conclusion}

We present an integrated, web-based platform that takes MR-Linac diffusion-weighted imaging from raw DICOM data to a structured, literature-grounded clinical interpretation by combining validated deep-learning-based distortion correction, denoising, and IVIM/ADC fitting with a retrieval-augmented interpretation agent that links its statements to specific passages in the source literature. In an independent evaluation of nine longitudinal glioblastoma cases, a medical physicist and a physician rated the agent-generated reports as clinically usable, with an overall score of $4.65\pm0.80$ on a 1--5 scale and the highest ratings for clinical utility. The same evaluation identified the system's remaining limitations (primarily inferred acquisition metadata and the use of imperfectly matched supporting evidence), and localized them to specific, correctable statements, illustrating how the platform's traceable design can make LLM-based reasoning auditable in radiation oncology. By integrating quantitative processing and literature-grounded interpretation within a single verifiable platform, this work represents a step toward broader use of multiparametric IVIM for treatment response monitoring beyond specialized centers.

\section*{Acknowledgments}
This research was supported by the National Institutes of Health (NIH) Grants R01 EB034691, R01 CA240808, R01 CA258987, and R01 CA280135.

\section*{Ethics Approval}
All datasets were retrospectively collected from an approved study at the UT Southwestern Medical Center, under an umbrella IRB protocol 082013-008 (Improving radiation treatment quality and safety by retrospective data analysis). This is a retrospective analysis study and not a clinical trial. No clinical trial ID number is available.

\section*{Conflict of Interest}
The authors have no relevant conflicts of interest to disclose.

\section*{Data Availability Statement}
The data cannot be made publicly available upon publication because they contain sensitive personal information. The data that support the findings of this study are available upon reasonable request from the authors.

\bibliographystyle{unsrtnat}
\bibliography{ref}

\end{document}